\documentclass{article}

\usepackage{amssymb}
\usepackage{amsmath}

\newtheorem{theorem}{Theorem}

\newtheorem{definition}[theorem]{Definition}

\newtheorem{proposition}[theorem]{Proposition}

\begin{document}

\title{\textbf{$\mathbb{Z}_{2}$-gradings of Clifford algebras and 
multivector structures}}
\author{\textbf{R. A. Mosna}$^{(1,2)}$\thanks{E-mail address: mosna@ifi.unicamp.br}
\textbf{, D. Miralles}$^{(3)}$\thanks{Current address: Departament de Comunicacions
i Teoria del Senyal, Universitat Ramon Llull, CP 08022, Barcelona, Catalonia. E-mail address:
davidme@salleurl.edu}
\textbf{\ }and 
\textbf{J. Vaz Jr}$^{(2)}$\thanks{E-mail address: vaz@ime.unicamp.br}\\
{\small (1) Instituto de F\'{\i}sica Gleb Wataghin, CPG, Universidade Estadual de Campinas,}\\
{\small CP 6165, 13083-970, Campinas, SP, Brazil.}\\
{\small (2) Departamento de Matem\'{a}tica Aplicada, Universidade Estadual de Campinas,}\\
{\small CP 6065, 13081-970, Campinas, SP, Brazil.}\\
{\small (3) Departament de F\'{\i}sica Fonamental, Universitat de Barcelona,}\\
{\small CP 08028, Barcelona, Catalonia.}}
\maketitle

\begin{abstract}
Let $\mathcal{C}\ell(V,g)$ be the real Clifford algebra associated to the real
vector space $V$, endowed with a nondegenerate metric $g$. In this paper, we
study the class of $\mathbb{Z}_{2}$-gradings of $\mathcal{C}\ell(V,g)$ which
are somehow compatible with the multivector structure of the Grassmann algebra
over $V$. A complete characterization for such $\mathbb{Z}_{2}$-gradings is
obtained by classifying all the even subalgebras coming from them. An
expression relating such subalgebras to the usual even part of $\mathcal{C}\ell(V,g)$ 
is also obtained. Finally, we employ this framework to define
spinor spaces, and to parametrize all the possible signature changes on
$\mathcal{C}\ell(V,g)$ by $\mathbb{Z}_{2}$-gradings of this algebra.

\end{abstract}

\section{Introduction}

Clifford algebras have long been an important tool in the interplay among
geometry, algebra and physics. The development of the theory of spinor
structures, with applications in field and string theories, and the study of
Dirac operators, with applications in geometry and topology, are examples of
this general setting. These algebras carry $\mathbb{Z}_{2}$-graded structures
which play a major role in such developments. For example, the usual
$\mathbb{Z}_{2}$-graded structure of the Clifford bundle over a Riemannian
manifold may be used to construct models of supersymmetric quantum mechanics,
which have unveiled deep connections between field theory and geometry
\cite{Frohlich98}. Also, in Hestenes's approach to Dirac theory
\cite{Hestenes67,Hestenes75,Hestenes95}, the usual $\mathbb{Z}_{2}$-grading of
the spacetime algebra is extensively employed to represent spinors by even
elements of this algebra (such an approach has a natural generalization for
arbitrary Clifford algebras \cite{Dimakis89}).

Let $V$ be a finite-dimensional real vector space endowed with a metric $g$.
By that we mean that $g:V\times V\rightarrow\mathbb{R}$ is a bilinear,
symmetric and nondegenerate map. Let $\mathcal{C}\ell(V,g)$ be the real
Clifford algebra associated to $(V,g)$. As a \textit{vector space}, 
$\mathcal{C}\ell(V,g)$ is naturally $\mathbb{Z}$-graded by the multivector
structure inherited from the Grassmann algebra $\Lambda(V)$ over $V$. 
This is the usual Chevalley construction (see eq. (\ref{Chevalley map})).
However, as an \textit{algebra}, $\mathcal{C}\ell(V,g)$ is not $\mathbb{Z}$-graded,
but only $\mathbb{Z}_{2}$-graded and, in general, such $\mathbb{Z}_{2}$-gradings
do not have to preserve in any sense the homogeneous subspaces of 
$\Lambda(V)\cong\mathcal{C}\ell(V,g)$ ($\cong$ denotes \textit{linear}
isomorphism in this expression).

In this paper, we study the class of $\mathbb{Z}_{2}$-gradings of
$\mathcal{C}\ell(V,g)$ which are somehow compatible with the usual multivector
structure of $\Lambda(V)$ (see definition \ref{def Z2 pres multivec struc}).
In section \ref{section Z2-gradings}, we completely characterize such
$\mathbb{Z}_{2}$-gradings by classifying all the even subalgebras coming from
them. Also, a formula relating such arbitrary even subalgebras to the usual
even part of $\mathcal{C}\ell(V,g)$ is obtained.

In the next section, some preliminary applications are considered. We start by
discussing the possibility of employing these arbitrary $\mathbb{Z}_{2}$-gradings 
to define spinor spaces, as in \cite{Dimakis89,Mosna02}. After
that, we consider the problem of signature change in an arbitrary Clifford
algebra. There are various situations in theoretical physics where changing
the signature of a given space is an useful tool, as in Euclidean formulations
of field theories, in the theory of instantons, in finite temperature field
theory and in lattice gauge theory. In \cite{Lounesto97} and \cite{Miralles01}, 
the authors discuss the specific signature changes $(1,3)\rightarrow(3,1)$
and $(1,3)\rightarrow(4,0)$ inside the spacetime algebra (in the last case,
the corresponding signature change map is used to study the Dirac equation,
and self-dual/anti-self-dual solutions of gauge fields). In section
\ref{section signature change}, we generalize such approaches in order to
obtain completely arbitrary signature change maps in Clifford algebras of any
dimension. As in the aforementioned works, our method is \textit{purely algebraic}, 
and is implemented by a deformation of the algebraic structure
underlying the theory. More specifically, the $\mathbb{Z}_{2}$-gradings
discussed above are employed to deform the original Clifford product, thereby
``simulating'' the product properties of the signature changed space. 
As a result, we parametrize all the possible signature changes on 
$\mathcal{C}\ell(V,g)$ by $\mathbb{Z}_{2}$-gradings of this algebra. This opens 
the possibility of applying this formalism to higher dimensional physical theories.

The concept of $\mathbb{Z}_{2}$-graded structures has numerous applications in
mathematical physics (as in supersymmetry, supergeometry, etc.). It is then
reasonable to expect that the present work may find other applications
besides the ones considered here and outlined above.

\subsection{Algebraic preliminaries and notation}

\label{section notation}

A vector space $W$ is said to be graded by an Abelian group $G$ if it is
expressible as a direct sum $W={\bigoplus\nolimits_{i}}W_{i}$ of subspaces
labelled by elements $i\in G$ (we refer the reader to appendix A of
\cite{BennTucker87} for a general review of algebraic concepts). Here 
we consider only the cases when $G$ is
given by $\mathbb{Z}$ or $\mathbb{Z}_{2}$. Then, the elements of $W_{i}$ are
called homogeneous of degree $i$ and we define $\deg(w)=i$ if $w\in W_{i}$.
Let $\mathcal{A}$ be an algebra which, for the purposes of this paper, can
always be considered as a finite-dimensional associative algebra with unit,
over $\mathbb{R}$ or $\mathbb{C}$. We say that $\mathcal{A}$ is graded by $G$
if (a) its underlying vector space is a $G$-graded vector space and (b) its
product satisfies $\deg(ab)=\deg(a)+\deg(b)$.

Let $V$ be an $n$-dimensional real vector space. Then, the tensor algebra
$T(V)={\bigoplus\nolimits_{k=0}^{\infty}}T^{k}(V)$ over $V$ is an example of
a $\mathbb{Z}$-graded algebra. We denote the space of antisymmetric
$k$-tensors by $\Lambda^{k}(V)$. The elements of this space will be called
$k$-vectors. Let 
$V^{\wedge}={\bigoplus\nolimits_{k=0}^{n}}\Lambda^{k}(V)$ denote the 
$2^{n}$-dimensional real vector space of multivectors over $V$. Using the 
natural embeddings of $\mathbb{R}$ and $V$ in $V^{\wedge }$, we identify 
$\Lambda ^{0}(V)$ with $\mathbb{R}$ and $\Lambda^{1}(V)$ with $V$. When endowed 
with the exterior product $\wedge$, the vector space $V^{\wedge}$ becomes 
the so called Grassmann algebra $\Lambda(V)=(V^{\wedge},\wedge)$ over $V$. 
We note that $\Lambda(V)={\bigoplus\nolimits_{k=0}^{n}}\Lambda^{k}(V)$ is 
another example of a $\mathbb{Z}$-graded algebra, with a $\mathbb{Z}$-graded 
structure inherited from the usual $\mathbb{Z}$-grading of $T(V)$. It is 
important to note that such $\mathbb{Z}$-grading for $\Lambda(V)$ is by no means unique 
\cite{Oziewicz97,FauserAblamowicz00}. Nevertheless, suppose one wants to identify
$V$ with the tangent space (at a certain point) of a spacetime $M$. Then, in the
context of this usual grading, one can interpret elements of $\Lambda^{0}(V)$ 
as scalars, elements of $\Lambda^{1}(V)$ as tangent vectors of $M$ and so on. 
In this paper, we always consider the multivector structure coming from such usual 
$\mathbb{Z}$-grading of $\Lambda(V)$ (more discussion along these lines can be 
found in \cite{Mosna02}).

We denote the projection of a multivector $a=a_{0}+a_{1}+\cdots+a_{n}$, with
$a_{k}\in\Lambda^{k}(V)$, on its $p$-vector part by $\langle a\rangle_{p}:=a_{p}$. 
The parity operator $(\cdot)^{\wedge}$ is defined as the algebra
automorphism generated by the expression $\hat{v}=-v$ on vectors $v\in V$. The
reversion $(\cdot)^{\sim}$ is the algebra anti-automorphism generated by the
expression $\tilde{v}=v$ on vectors $v\in V.$ It follows that $\hat{a}=(-1)^{k}a$ 
and $\tilde{a}=(-1)^{[k/2]}a$ if $a\in\Lambda^{k}(V)$, where
$[m]$ denotes the integer part of $m$. When $V$ is endowed with a metric $g$, it 
is possible to extend (in a non-unique way) $g$ to all of $V^{\wedge}$.
Given $a=u_{1}\wedge\cdots\wedge u_{k}$ and $b=v_{1}\wedge\cdots\wedge v_{l}$
with $u_{i},v_{j}\in V$, the expressions $g(a,b)=\det(g(u_{i},v_{j})) $, if
$k=l$, and $g(a,b)=0$, if $k\neq l$, provides one such extension. Also, the
left ($\lrcorner$) and right ($\llcorner$) contractions on the Grassmann
algebra are respectively defined by $g(a\lrcorner b,c)=g(b,\tilde{a}\wedge c)$
and $g(b\llcorner a,c)=g(b,c\wedge\tilde{a})$, with $a,b,c\in\Lambda(V)$.

The Clifford product between a vector $v\in V$ and a multivector $a$ in
$V^{\wedge}$ is given by $va=v\wedge a+v\lrcorner a$. This is extended by
linearity and associativity to all of $V^{\wedge}$. The resulting algebra is
the so called Clifford algebra $\mathcal{C}\ell(V,g)$. Note that, although the
underlying vector space of $\mathcal{C}\ell(V,g)$ (i.e., $V^{\wedge}$) 
is $\mathbb{Z}$-graded, $\mathcal{C}\ell(V,g)$ is not a $\mathbb{Z}$-graded 
algebra as, for example, the Clifford product between two 1-vectors
is a sum of elements of degrees $0$ and $2$. Nevertheless, there are
(infinite) $\mathbb{Z}_{2}$-gradings which are compatible with the Clifford
product structure. For instance, the usual $\mathbb{Z}_{2}$-grading of
$\mathcal{C}\ell(V,g)$ is given by 
$\mathcal{C}\ell^{+}(V,g)\oplus\mathcal{C}\ell^{-}(V,g)$ where 
$\mathcal{C}\ell^{+}(V,g)={\bigoplus\nolimits_{k\,even}}\Lambda^{k}(V)$ and 
$\mathcal{C}\ell^{-}(V,g)={\bigoplus\nolimits_{k\,odd}}\Lambda^{k}(V)$. When 
the metric $g$ has signature $(p,q)$, we will also denote the real vector space 
$V$ endowed with $g$ by $\mathbb{R}^{p,q}$. In this case, the real Clifford 
algebra $\mathcal{C}\ell(V,g)$ over $V$ will be denoted by 
$\mathcal{C}\ell_{p,q}(\mathbb{R})$ or $\mathcal{C}\ell_{p,q}$. We adopt the definition 
$\mathcal{C}\ell_{p,q}(\mathbb{C})=\mathcal{C}\ell_{p,q}(\mathbb{R})\otimes\mathbb{C} $ for
the complexified Clifford algebra (of course, all the $\mathcal{C}\ell_{p,q}(\mathbb{C})$ 
with fixed $p+q$ are isomorphic as complex algebras). Note that given 1-vectors 
$x,y\in\mathbb{R}^{p,q}$, we have $2g(x,y)=xy+yx$. In particular, an orthonormal 
basis $\{e_{i}\}$ of $\mathbb{R}^{p,q}$ yields $e_{i}e_{j}+e_{j}e_{i}=2g_{ij}$, 
where $g_{ij}=g(e_{i},e_{j})$. In the following, we denote by $\mathcal{M}(m,\mathbb{K})$ 
the space of $m\times m$ matrices over $\mathbb{K}$, where 
$\mathbb{K}=\mathbb{R}$, $\mathbb{C}$ or $\mathbb{H}$.

We observe that there are other ways of defining Clifford and Grassmann algebras 
(see, for example, chapter 14 of \cite{Lounesto97} and chapters 1 and 2 of 
\cite{BennTucker87}). In the definitions adopted here, both the Grassmann and 
the Clifford algebras are defined on the same underlying vector space $V^{\wedge}$. 
This will be particularly useful in section \ref{section signature change}, where 
we consider various Clifford products defined, at the same time, on $V^{\wedge}$.

It is well known that real Clifford algebras exhibit an 8-fold periodicity and can be 
classified by 
$\mathcal{C}\ell_{p,q}(\mathbb{R})\cong\mathcal{M}(m,\mathbb{R})\otimes\mathcal{A}$, 
where $\mathcal{A}$ is given by table \ref{table real CA} and $m$ is fixed 
by $m^{2}\dim_{\mathbb{R}}\mathcal{A}=2^{n},$ with $n=p+q$.
\begin{table}[htbp]
\centering
\begin{tabular}{|c|c|c|c|c|c|c|c|c|}
\hline
$p-q$ $(\operatorname{mod}8)$ & $0$ & $1$ & $2$ & $3$ & $4$ & $5$ & $6$ & $7$\\
\hline
$\mathcal{A}$ & $\mathbb{R}$ & $\mathbb{R}\oplus \mathbb{R}$ & $\mathbb{R}$ &
$\mathbb{C}$ & $\mathbb{H}$ & $\mathbb{H}\oplus \mathbb{H}$ & $\mathbb{H}$ & $\mathbb{C}$\\
\hline
\end{tabular}
\caption{Classification of real Clifford algebras 
$\mathcal{C}\ell_{p,q}(\mathbb{R})\protect\cong\mathcal{M}(m,\mathbb{R})\otimes\mathcal{A}$, 
where $m^{2}\dim_{\mathbb{R}}\mathcal{A}=2^{n}$ and $n=p+q$. \label{table real CA}}
\end{table}

The usual even subalgebras $\mathcal{C}\ell_{p,q}^{+}(\mathbb{R})$ can be shown to satisfy
\begin{equation}
\mathcal{C}\ell_{p,q}^{+}(\mathbb{R})\cong\mathcal{C}\ell_{q,p-1}(\mathbb{R})\cong
\mathcal{C}\ell_{p,q-1}(\mathbb{R})\cong\mathcal{C}\ell_{q,p}^{+}(\mathbb{R}). 
\label{usual even part}
\end{equation}
In this way, their classification follows from table \ref{table real CA}, as
table \ref{table usual even part of real CA} shows.
\begin{table}[htbp]
\centering
\begin{tabular}{|c|c|c|c|c|c|c|c|c|}
\hline
$p-q$ $(\operatorname{mod}8)$ & $0$ & $1$ & $2$ & $3$ & $4$ & $5$ & $6$ & $7$\\
\hline
$\mathcal{B}$ & $\mathbb{R}\oplus \mathbb{R}$ & $\mathbb{R}$ & $\mathbb{C}$ &
$\mathbb{H}$ & $\mathbb{H}\oplus \mathbb{H}$ & $\mathbb{H}$ & $\mathbb{C}$ & $\mathbb{R}$\\
\hline
\end{tabular}
\caption{Usual even parts of real Clifford algebras 
$\mathcal{C}\ell_{p,q}^{+}(\mathbb{R})\protect\cong\mathcal{M}(m,\mathbb{R})\otimes\mathcal{B}$, 
where $m^{2}\dim_{\mathbb{R}}\mathcal{B}=2^{n-1}$ and $n=p+q$.
\label{table usual even part of real CA}}
\end{table}

The classification of the complex Clifford algebras is simpler, as table
\ref{table complex CA} shows (in this table, $\cong$ denotes isomorphism of
complex algebras).
\begin{table}[htbp]
\centering
\begin{tabular}{|c|c|}
\hline
even dimension & $\mathcal{C}\ell_{2k}(\mathbb{C})\cong\mathcal{M}(2^{k},\mathbb{C)}$\\
\hline
odd dimension & 
$\mathcal{C}\ell_{2k+1}(\mathbb{C})\cong\mathcal{M}(2^{k},\mathbb{C)}\oplus\mathcal{M}(2^{k},\mathbb{C)}$\\
\hline
\end{tabular}
\caption{Classification of complex Clifford algebras.
\label{table complex CA}}
\end{table}

Clifford algebras may also be characterized by their universal property, in
the sense of the well known theorem below.

\begin{theorem}
Let $V$ be a finite dimensional real vector space endowed with a nondegenerate
metric $g$. Let $\mathcal{A}$ be a real associative algebra with unit
$1_{\mathcal{A}}$. Given a linear map $\Gamma:V\rightarrow\mathcal{A}$ such
that $(\Gamma(v))^{2}=g(v,v)1_{\mathcal{A}}$, there exists a unique
homomorphism $\bar{\Gamma}:\mathcal{C}\ell(V,g)\rightarrow\mathcal{A}$ such
that $\bar{\Gamma}|_{V}=\Gamma$.\label{fundamental theorem}
\end{theorem}

A map $\Gamma$ as in the above theorem will be called a \emph{Clifford map}
for the pair $(V,g)$. An important example is given by the Clifford map
$\Gamma:V\rightarrow End(V^{\wedge})$, defined by
\begin{equation}
\Gamma(v)=v\wedge+v\lrcorner \ , \label{Chevalley map}
\end{equation}
which implements the well known Chevalley identification of $\mathcal{C}\ell(V,g)$ 
with a subalgebra of $End(V^{\wedge})$.

\section{$\mathbb{Z}_{2}$-gradings of Clifford Algebras}

\label{section Z2-gradings}

Abusing language, we will denote an arbitrary $\mathbb{Z}_{2}$-grading of 
$\mathcal{C}\ell(V,g)$ simply by 
$\mathcal{C}\ell(V,g)=\mathcal{C}\ell_{0}\oplus\mathcal{C}\ell_{1}$. 
In this way, (the vector space structure of) $\mathcal{C}\ell(V,g)$ is given by 
a direct sum of subspaces $\mathcal{C}\ell_{i}$, $i=0,1$, which satisfy
\begin{equation}
\mathcal{C}\ell_{i}\mathcal{C}\ell_{j}\subseteq\mathcal{C}\ell_{i+j(\operatorname{mod}2)}.
\label{Z2grading rule}
\end{equation}
Of course, $\mathcal{C}\ell_{0}$ is then a subalgebra of $\mathcal{C}\ell(V,g)$. 
To each such a decomposition we have an associated vector space
automorphism $\alpha:\mathcal{C}\ell(V,g)\rightarrow\mathcal{C}\ell(V,g)$
defined by $\alpha|_{\mathcal{C}\ell_{i}}=(-1)^{i}$id$_{\mathcal{C}\ell_{i}}$
(where id$_{W}$ denotes the identity map on the space $W$). The projections
$\pi_{i}$ on $\mathcal{C}\ell_{i}$ are given by $\pi_{i}(a)=\frac{a+(-1)^{i}\alpha(a)}{2}$. 
We also denote $\pi_{i}(a)=a_{i}$. Note that $\alpha$ is an algebra isomorphism, 
for given $a,b\in\mathcal{C}\ell(V,g)$, we have 
$\alpha(ab)=\alpha({\textstyle\sum_{ij}}a_{i}b_{j})=
{\textstyle\sum_{ij}}\alpha(a_{i}b_{j})=
{\textstyle\sum_{ij}}(-1)^{i+j}a_{i}b_{j}=
{\textstyle\sum_{i}}(-1)^{i}a_{i}{\textstyle\sum_{j}}(-1)^{j}b_{j}=
\alpha(a)\alpha(b)$. For the usual $\mathbb{Z}_{2}$-grading, where 
$\mathcal{C}\ell_{0}=\mathcal{C}\ell_{p,q}^{+}(\mathbb{R})$ and 
$\mathcal{C}\ell_{1}=\mathcal{C}\ell_{p,q}^{-}(\mathbb{R})$, 
the grading automorphism is simply given by $(\cdot)^{\wedge}$
(see section \ref{section notation}). 

Given a $\mathbb{Z}_{2}$-grading of $\mathcal{C}\ell(V,g)$ as above, 
we refer to $\mathcal{C}\ell_{0}$ and $\mathcal{C}\ell_{1}$ as the 
$\alpha$\textit{-even} and $\alpha$\textit{-odd} parts of $\mathcal{C}\ell(V,g)$.
Also, an element belonging to $\mathcal{C}\ell_{0}$ ($\mathcal{C}\ell_{1}$) will 
be called $\alpha$-even ($\alpha$-odd).

We observe that the scalar $1\in\Lambda^{0}(V)$ is always $\alpha$-even.
Indeed, let us write $1=e+o$, where $e=\pi_{0}(1)$ and $o=\pi_{1}(1)$.
Left-multiplying this equation by $e$ yields $e=e^{2}+eo$. As $e$ and $e^{2}$
are $\alpha$-even and $eo$ is $\alpha$-odd, we must have $eo=0$. Then,
right-multiplying $1=e+o$ by $o$ yields $o=o^{2}$. As $o$ is $\alpha$-odd and
$o^{2}$ is $\alpha$-even, we thus have $o=0$.

Let us now address the central point of the present paper. 
In the general $\mathbb{Z}_{2}$-gradings introduced so far, the even and odd
projections do not have to preserve the multivector structure of $\Lambda(V)$
(see example later). In other words, it is possible that the even or odd part
of a $k$-vector comprises an inhomogeneous combination of elements of
different degrees.

\begin{proposition}
Let $\mathcal{C}\ell(V,g)=\mathcal{C}\ell_{0}\oplus\mathcal{C}\ell_{1}$ be a
$\mathbb{Z}_{2}$-grading with grading automorphism $\alpha$. The following are 
equivalent:\footnote{In the following, we make the usual identification of $V$
with $\Lambda^{1}(V)$ (as mentioned in section \ref{section notation}). In
particular, items (ii) and (iv) of proposition \ref{proposition pres multivec struc} 
can be written, in a more precise way, as $\pi_{i}(\Lambda^{1}(V))\subseteq\Lambda^{1}(V)$ 
and $\alpha(\Lambda ^{1}(V))\subseteq\Lambda^{1}(V),$ respectively.}

(i) The projections $\pi_{i}$, $i=0,1$, preserve each $\Lambda^{k}(V)$, $k=1,\dots,n$;

(ii) $\pi_{i}(V)\subseteq V$, $i=0,1$;

(iii) $\alpha$ preserves each $\Lambda^{k}(V)$, $k=1,\dots,n$;

(iv) $\alpha(V)\subseteq V$.
\label{proposition pres multivec struc}
\end{proposition}

\noindent\textbf{Proof.:} It follows from the definition of $\pi_{i}$ that
$\pi_{i}(\Lambda^{k}(V))\subseteq\Lambda^{k}(V)$ if, and only if,
$\alpha(\Lambda^{k}(V))\subseteq\Lambda^{k}(V)$. Thus (i)$\Leftrightarrow$(iii) 
and (ii)$\Leftrightarrow$(iv). Of course, (iii)$\Rightarrow$(iv).
Conversely, if we assume (iv), then $\alpha|_{V}$ (see footnote 1) is an isometry. 
Indeed, given $x,y\in V$, 
$2g(\alpha(x),\alpha(y))=\alpha(x)\alpha(y)+\alpha(y)\alpha(x)=\alpha(xy+yx)=2g(x,y)$, 
since $\alpha$ is an algebra isomorphism and $\alpha(1)=1$ (as we mentioned earlier). 
If $\{e_{i}\}$ is an orthonormal basis of $V$, every element $a\in\Lambda^{k}(V)$
can be written as a linear combination of terms like 
$e_{i_{1}}\wedge\cdots\wedge e_{i_{k}}=e_{i_{1}}\cdots e_{i_{k}}$. As $\alpha|_{V}$ 
is an isometry, $\{\alpha(e_{i})\}$ is also an orthonormal basis of $V$ and thus
$\alpha(e_{i_{1}}\cdots e_{i_{k}})=\alpha(e_{i_{1}})\cdots\alpha(e_{i_{k}})=
\alpha(e_{i_{1}})\wedge\cdots\wedge\alpha(e_{i_{k}})\in\Lambda^{k}(V)$. It
follows that $\alpha(a)\in\Lambda^{k}(V)$, establishing (iv)$\Rightarrow$(iii).$\Box$

\begin{definition}
A $\mathbb{Z}_{2}$-grading fulfilling one (and hence all) of the conditions
above will be said to preserve the multivector structure of $\Lambda(V)$.
\label{def Z2 pres multivec struc}
\end{definition}

For this class of $\mathbb{Z}_{2}$-gradings, we have the following proposition.

\begin{proposition}
Let $\mathcal{C}\ell(V,g)=\mathcal{C}\ell_{0}\oplus\mathcal{C}\ell_{1}$ be a
$\mathbb{Z}_{2}$-grading preserving the multivector structure of $\Lambda(V)$
and define $V_{i}:=V\cap\mathcal{C}\ell_{i}$, $i=0,1$, i.e., $V_{0}$ ($V_{1}$)
is the space of $\alpha$-even ($\alpha$-odd) 1-vectors (see footnote 1). Then 
$V=V_{0}\oplus V_{1}$, with $V_{0}=V_{1}{}^{\perp}$ (and $V_{1}=V_{0}{}^{\perp}$). 
It follows that each subspace $V_{i}$ is nondegenerate (i.e. $g$ restricted 
to $V_{i}$ is nondegenerate).\label{prop V=V0+V1}
\end{proposition}

\noindent\textbf{Proof.:} By assumption, each projection $\pi_{i}$ preserves
$V$. This immediately induces a $\mathbb{Z}_{2}$-grading for the vector space
$V$, so that $V=V_{0}\oplus V_{1}$. Moreover, such a decomposition is
orthogonal. Indeed, given $x\in V_{0}$ and $y\in V_{1}$, we have
$xy+yx=2g(x,y)$. As the left hand side belongs to $\mathcal{C}\ell_{1}$ and
the right hand side to $\mathcal{C}\ell_{0}$, we must have $g(x,y)=0$. Thus
$V_{0}\perp V_{1}$ and, in particular, $V_{0}\subseteq V_{1}{}^{\perp}$. By
counting dimensions, we finally have $V_{0}=V_{1}{}^{\perp}$ (for 
$\dim(V_{0})+\dim(V_{1})=n$ and $\dim(V_{1})+\dim(V_{1}{}^{\perp})=n$).$\Box
\vspace{10pt}$

Let the metric $g$ have signature $(p,q)$, with $p+q=n$, and let us denote the
vector space $V$, endowed with $g$, by $\mathbb{R}^{p,q}$ (as in section
\ref{section notation}). By proposition \ref{prop V=V0+V1}, we can
choose orthonormal basis $\mathcal{B}_{0}=\{v_{1},\dots,v_{a}\}$ and
$\mathcal{B}_{1}=\{v_{a+1},\dots,v_{a+b}\}$ of $V_{0}$ and $V_{1}$
respectively. Then, $\{v_{1},\dots,v_{a+b}\}$ is an orthonormal basis for
$\mathbb{R}^{p,q}$ and thus 
$\{1,v_{i_{1}}\cdots v_{i_{k}}:1\leq i_{1}\leq\cdots\leq i_{k}\leq n$, 
$k=1,\dots,n\}$ is an orthonormal basis for $\mathcal{C}\ell_{p,q}(\mathbb{R})$. 
It follows from eq. (\ref{Z2grading rule}) that $\mathcal{C}\ell_{0}$ is 
generated (as an algebra) by elements of the form:
\begin{equation}
\begin{tabular}{ll}
(i) $v_{i}$, & with $i\leq a$,\\
(ii) $v_{i}v_{j}$, & with $i,j>a$.
\end{tabular}
\label{generators of Cl0}
\end{equation}
This leads to a straightforward characterization of $\mathcal{C}\ell_{0}$. Let
$p_{i}$ $(q_{i})$ be the number of elements in $\mathcal{B}_{i}$ squaring
to $+1$ $(-1)$. As we are working within $\mathcal{C}\ell_{p,q}(\mathbb{R})$, we have 
$p=p_{0}+p_{1}$ and $q=q_{0}+q_{1}.$ Then, eq. (\ref{generators of Cl0}) implies 
$\mathcal{C}\ell_{0}\cong\mathcal{C}\ell_{p_{0},q_{0}}\otimes\mathcal{C}\ell_{p_{1},q_{1}}^{+}$, 
where the tensor product is over $\mathbb{R}$ and comes from the fact that the elements in (i)
and (ii) commute. We summarize this result in the following proposition.

\begin{proposition}
Let $\mathcal{C}\ell_{p,q}(\mathbb{R})=\mathcal{C}\ell_{0}\oplus\mathcal{C}\ell_{1}$ 
be a $\mathbb{Z}_{2}$-grading preserving the multivector structure of 
$\Lambda(\mathbb{R}^{p,q})$. Then
\begin{equation}
\mathcal{C}\ell_{0}\cong
\mathcal{C}\ell_{p_{0},q_{0}}\otimes\mathcal{C}\ell_{p-p_{0},q-q_{0}}^{+}, 
\label{general form of Cl0}
\end{equation}
where $p_{0}$ $(q_{0})$ is the number of $\alpha$-even elements of an
orthonormal basis of $\mathbb{R}^{p,q}$ squaring to $+1$ $(-1)$.
\label{proposition general form o Cl0}
\end{proposition}

Note that:

(i) if $p_{0}=q_{0}=0$, then eq. (\ref{general form of Cl0}) reduces to
$\mathcal{C}\ell_{0}=\mathcal{C}\ell_{p,q}^{+}(\mathbb{R})$, as expected. We
refer to this case as the \textit{usual} $\mathbb{Z}_{2}$-grading of
$\mathcal{C}\ell_{p,q}(\mathbb{R})$.

(ii) if $p_{0}=p$ and $q_{0}=q$, then $\mathcal{C}\ell_{0}=\mathcal{C}\ell_{p,q}(\mathbb{R})$ 
and $\mathcal{C}\ell_{1}=0$. Therefore, this case corresponds to the \textit{trivial} 
$\mathbb{Z}_{2}$-grading of $\mathcal{C}\ell_{p,q}(\mathbb{R})$. Moreover, this is the 
unique choice for $p_{0}$ and $q_{0}$ which yields the trivial $\mathbb{Z}_{2}$-grading.

It follows that every non-trivial $\mathbb{Z}_{2}$-grading, preserving the
multivector structure of $\Lambda(V)$, provides an invertible $\alpha$-odd
element $u$ (for example, any basis 1-vector in $V_{1}$ squaring to $\pm1$ can
be chosen for $u$). This can be used to construct the isomorphism of vector
spaces $\mathcal{C}\ell_{0}\rightarrow\mathcal{C}\ell_{1}$, $x\mapsto ux$.
Thus, the class of $\mathbb{Z}_{2}$-gradings considered here is such that
either
\begin{subequations}\label{dim Cl0}
\begin{align}
\textrm{(a)} \dim\mathcal{C}\ell_{0} &  =\dim\mathcal{C}\ell_{p,q}(\mathbb{R}) 
\textrm{ (trivial case), or }\label{dim Cl0 a}\\
\textrm{(b)} \dim\mathcal{C}\ell_{0} &  =\tfrac{1}{2}\dim\mathcal{C}\ell_{p,q}(\mathbb{R}).\label{dim Cl0 b}
\end{align}
\end{subequations}

At this point, it is interesting to consider some examples of $\mathbb{Z}_{2}$-gradings 
that do not preserve the multivector structure of $\Lambda(V)$. For
simplicity, let us momentarily regard the real Clifford algebra $\mathcal{C}\ell(V,g)$ 
as an algebra of $m\times m$ matrices, as in table \ref{table real CA}. Let us define 
$\mathcal{C}\ell_{0}$ and $\mathcal{C}\ell_{1}$ as, respectively, the spaces of 
matrices of the form
\[
\left(
\begin{array}{cc}
A & 0_{a\times b}\\
0_{b\times a} & B
\end{array}
\right)  \qquad\textrm{and}\qquad
\left(
\begin{array}{cc}
0_{a\times a} & C\\
D & 0_{b\times b}
\end{array}
\right) ,
\]
where $A$ and $B$ are square matrices of order $a$ and $b$, respectively, with
$a+b=m$. It is easy to see that this gives a $\mathbb{Z}_{2}$-grading for
$\mathcal{C}\ell(V,g)$ for any choice of $a$ and $b$. In particular, if $m>2$,
we can choose nonzero $a$ and $b$ such that $a\neq b$. As a result, we end up
with a nontrivial $\mathbb{Z}_{2}$-grading with 
$\dim\mathcal{C}\ell_{0}\neq\frac{1}{2}\dim\mathcal{C}\ell_{p,q}(\mathbb{R})$. 
It follows from eqs. (\ref{dim Cl0}) that such $\mathbb{Z}_{2}$-grading does 
not preserve the multivector structure of $\Lambda(V)$.

Let us now return to the study of the $\mathbb{Z}_{2}$-gradings preserving the
multivector structure of $\Lambda(\mathbb{R}^{p,q})$. The explicit formula for
$\mathcal{C}\ell_{0}$ in proposition \ref{proposition general form o Cl0} can
be used to obtain a complete classification for these objects. In fact, a
straightforward calculation (using the facts that 
$\mathbb{C}\otimes \mathbb{C}\cong \mathbb{C}\oplus \mathbb{C}$, 
$\mathbb{C}\otimes \mathbb{H}\cong \mathbb{C}\otimes\mathcal{M}(2,\mathbb{R})$ and
$\mathbb{H}\otimes \mathbb{H}\cong \mathcal{M}(4,\mathbb{R})$) shows that 
$\mathcal{C}\ell_{0}\cong\mathcal{M}(k,\mathbb{R})\otimes\mathcal{D}$, where $\mathcal{D}$
is given by table \ref{table Cl0} and $k$ is fixed by 
$k^{2}\dim_{\mathbb{R}}\mathcal{D}=2^{n-1}$, where $n=p+q$. It is interesting to note 
that this yields an overall 4-fold periodicity in terms of $p_{0}-q_{0}$.
\begin{table}[htbp]
\centering
\begin{tabular}{|c|c|c|c|c|c|c|c|c|}
\hline
$_{p_{0}-q_{0}}\diagdown^{p-q}$ & $0$ & $1$ & $2$ & $3$ & $4$ & $5$ & $6$ & $7$\\
\hline
$0$ & $\mathbb{R}\oplus \mathbb{R}$ & $\mathbb{R}$ & $\mathbb{C}$ & $\mathbb{H}$ &
$\mathbb{H}\oplus \mathbb{H}$ & $\mathbb{H}$ & $\mathbb{C}$ & $\mathbb{R}$\\
\hline
$1$ & $\mathbb{R}\oplus \mathbb{R}$ & 
$\mathbb{R}\oplus \mathbb{R}\oplus \mathbb{R}\oplus \mathbb{R}$ &
$\mathbb{R}\oplus \mathbb{R}$ & $\mathbb{C}\oplus \mathbb{C}$ & $\mathbb{H}\oplus \mathbb{H}$ &
$\mathbb{H}\oplus \mathbb{H}\oplus \mathbb{H}\oplus \mathbb{H}$ & 
$\mathbb{H}\oplus \mathbb{H}$ & $\mathbb{C}\oplus\mathbb{C}$\\
\hline
$2$ & $\mathbb{C}$ & $\mathbb{R}$ & $\mathbb{R}\oplus \mathbb{R}$ & $\mathbb{R}$ &
$\mathbb{C}$ & $\mathbb{H}$ & $\mathbb{H}\oplus \mathbb{H}$ & $\mathbb{H}$\\
\hline
$3$ & $\mathbb{C}$ & $\mathbb{C}\oplus \mathbb{C}$ & $\mathbb{C}$ & 
$\mathbb{C}\oplus \mathbb{C}$ & $\mathbb{C}$ & $\mathbb{C}\oplus \mathbb{C}$ & 
$\mathbb{C}$ & $\mathbb{C}\oplus\mathbb{C}$\\
\hline
\end{tabular}
\caption{Even subalgebras ($\mathcal{C}\ell_{0}$'s) associated to 
$\mathbb{Z}_{2}$-gradings preserving the multivector structure of 
$\Lambda(\mathbb{R}^{p,q})$. The table exhibits $\mathcal{D}$ in 
$\mathcal{C}\ell_{0}\protect\cong\mathcal{M}(k,\mathbb{R})\otimes\mathcal{D}$, 
where $k^{2}\dim_{\mathbb{R}}\mathcal{D}=2^{n-1}$ and $n=p+q$. Here, 
$p-q$ and $p_{0}-q_{0}$ should be considered $\operatorname{mod}8$ and 
$\operatorname{mod}4$ respectively. \label{table Cl0}}
\end{table}

We see from table \ref{table Cl0} that $\mathcal{C}\ell_{0}$ is not always a
Clifford algebra. For instance, when $p-q=1(\operatorname{mod}4)$ and
$p_{0}-q_{0}=1(\operatorname{mod}4)$, we have 
$\mathcal{C}\ell_{0}\cong\mathcal{M}(k,\mathbb{R})\otimes
(\mathbb{R}\oplus \mathbb{R}\oplus \mathbb{R}\oplus \mathbb{R})$, and
we know that no Clifford algebra has this form.

\section{Applications}

\label{section applications}

Now we consider some simple applications of the framework developed in the previous
section. In section \ref{section spinors}, we outline a possible use of the
$\mathbb{Z}_{2}$-gradings studied here to define spinor spaces, as in
\cite{Dimakis89} and \cite{Mosna02}. In section \ref{section signature change}, 
we analyze an algebraic method for changing the signature of arbitrary real
Clifford algebras, as advanced at the introduction.

\subsection{Spinor spaces}

\label{section spinors}

The identification of the even part of a Clifford algebra with a space of
spinors is mostly known in the context of Hestenes's formulation of Dirac
theory \cite{Hestenes67,Hestenes75,Hestenes95}. In such an approach, the state
of the electron is described by an operator spinor \cite{Figueiredo90}
$\Psi\in\mathcal{C}\ell_{1,3}^{+}(\mathbb{R})$ satisfying the so called
Dirac-Hestenes equation, $\partial\Psi e_{21}=m\Psi e_{0}$ (here $\{e_{\mu}\}$
is an orthonormal frame in Minkowski space, corresponding to a given observer,
and $\partial=e^{\mu}\partial_{\mu}$). We observe that the space of operator
spinors is more than a vector space, it is an \emph{algebra}. This leads,
among other things, to an elegant canonical decomposition for $\Psi$, which
generalizes the polar decomposition of complex numbers. The Dirac-Hestenes
equation is covariant under a change of frame/observer, for another choice
$\{e_{\mu}^{\prime}\}$ must be related to the old one by $e_{\mu}^{\prime}=Ue_{\mu}\tilde{U}$, 
with $U\in Spin_{+}(1,3)$, yielding 
$\partial\Psi^{\prime}e_{21}^{\prime}=m\Psi^{\prime}e_{0}^{\prime}$, where
$\Psi^{\prime}=\Psi\tilde{U}$. On the other hand, the usual (matrix) Dirac
equation is known to be covariant under a larger class of transformations, in
which the gamma matrices $\gamma_{\mu}$ are transformed by 
$\mathsf{S}\gamma_{\mu}\mathsf{S}^{-1}$, where $\mathsf{S}$ is an arbitrary unitary
matrix (this amounts to a change in the gamma matrix representation).

By considering this kind of transformations, it is possible to derive
multivector Dirac equations associated to a large class of gamma matrix
representations, including the standard, Majorana and chiral ones
\cite{Mosna02}. The resulting spinor spaces can be identified with even
subalgebras $\mathcal{C}\ell_{0}$ of the kind considered in the previous
section. Indeed, the generalized Dirac-Hestenes equation in this context reads
$\breve{\partial}\Psi\sigma+m\Psi u=0$, where $\Psi\in\mathcal{C}\ell_{0}$,
$\sigma$ and $u$ are any commuting $\alpha$-even and $\alpha$-odd elements,
respectively, satisfying $\sigma^{2}=-1$ and $u^{2}=1$, and 
$\breve{\partial}\Psi:=\pi_{0}(\partial)\Psi u+\pi_{1}(\partial)\Psi$ 
(see \cite{Mosna02} for details). It follows that the resulting operator spinor 
spaces for the Dirac theory are isomorphic to either $\mathcal{M}(2,\mathbb{C})$ or
$\mathbb{H}\oplus \mathbb{H}$. This method gives rise to a generalized spinor map,
relating algebraic and operator spinors, which was used by us \cite{Mosna02}
to rederive certain quaternionic models of (the usual) quantum mechanics and
to provide a natural way to obtain gamma matrix representations in terms of
the enhanced $\mathbb{H}$-general linear group 
$GL(2,\mathbb{H})\cdot\mathbb{H}^{\ast}$ \cite{Harvey90}.

Let us now briefly consider more general Clifford algebras than 
$\mathcal{C}\ell_{1,3}(\mathbb{R})$. As it was shown by A. Dimakis, it is 
always possible to represent a given $\mathcal{C}\ell_{p,q}(\mathbb{R})$ in itself, 
with a corresponding spinor space isomorphic to a subalgebra of the original algebra.
This is done in \cite{Dimakis89}, where such a subalgebra is obtained by
taking the even part of successive $\mathbb{Z}_{2}$-gradings of 
$\mathcal{C}\ell_{p,q}(\mathbb{R})$. Moreover, this subalgebra is a real Clifford algebra
by itself. Let us now outline a slight generalization of this procedure, in
which the corresponding $\mathbb{Z}_{2}$-gradings are given as in the previous
section. As we have seen, the resulting even subalgebra is not necessarily a
real Clifford algebra in this case.

First of all, we note that the even subalgebra $\mathcal{C}\ell_{0}$ is in
general too large to be taken as the space of spinors, which is classically
given by a minimal one-sided ideal $\mathcal{I}$ in 
$\mathcal{C}\ell_{p,q}(\mathbb{R})$ \cite{Chevalley54}. Indeed, we have shown 
in the previous section that, for the (non-trivial) $\mathbb{Z}_{2}$-gradings 
considered here, $\dim\mathcal{C}\ell_{0}=\frac{1}{2}\dim\mathcal{C}\ell_{p,q}(\mathbb{R})$.
Thus, $\mathcal{I}$ and $\mathcal{C}\ell_{0}$ have the same dimension only for
Clifford algebras isomorphic to $2\times2$ matrices, i.e., for 
$\mathcal{C}\ell_{2,0}(\mathbb{R})\cong\mathcal{M}(2,\mathbb{R})$, 
$\mathcal{C}\ell_{3,0}(\mathbb{R})\cong\mathcal{M}(2,\mathbb{C})$ and 
$\mathcal{C}\ell_{1,3}(\mathbb{R})\cong\mathcal{M}(2,\mathbb{H})$ (modulo isomorphisms)
(see also section 10.8 of \cite{Lounesto97}). 
As we have already mentioned, the case $\mathcal{C}\ell_{1,3}(\mathbb{R})$ was
analyzed in \cite{Mosna02}. On the other hand, for the Clifford algebra
$\mathcal{C}\ell_{3,0}(\mathbb{R})$, which is related to Pauli theory in the
same way as $C\ell_{1,3}(\mathbb{R})$ is related to Dirac theory, our method
leads to spinor spaces isomorphic to $\mathbb{H}$, $\mathcal{M}(2,\mathbb{R})$
or $\mathbb{C}\oplus\mathbb{C}$ (see table \ref{table Cl0}). Note that, as 
$\mathbb{C}\oplus \mathbb{C}$ is not a real Clifford algebra, this case is not
given by Dimakis's method. A study of Pauli equation along the lines of
\cite{Mosna02} would then result in three different (i.e. non-isomorphic)
corresponding spinor algebras for this case.

For higher dimensional Clifford algebras, we can successively take
$\mathcal{C}\ell_{0}$, $\mathcal{C}\ell_{00}=(\mathcal{C}\ell_{0})_{0}$ and so
one, having in mind that the prescription given by eq.
(\ref{general form of Cl0}) only works when we have a real Clifford algebra
involved. In the other cases, one might consider further generalizations of
eq. (\ref{general form of Cl0}), like 
$\mathcal{C}\ell_{p_{0},q_{0}}^{+}\otimes\mathcal{C}\ell_{p-p_{0},q-q_{0}}^{+}$ or 
$\mathcal{C}\ell_{p_{0},q_{0}}\otimes\mathcal{C}\ell_{p-p_{0},q-q_{0}}^{++}$ for example.

\subsection{Signature change in Clifford algebras}

\label{section signature change}

Let us now associate to each $\mathbb{Z}_{2}$-grading 
$\mathcal{C}\ell(V,g)=\mathcal{C}\ell_{0}\oplus\mathcal{C}\ell_{1}$, 
with corresponding grading automorphism $\alpha$, 
the linear map $\Gamma_{\alpha}:V\rightarrow End(V^{\wedge})$
given by (cf eq. (\ref{Chevalley map})):
\[
\Gamma_{\alpha}(v)=v\wedge+\alpha(v)\lrcorner\;.
\]

\begin{proposition}
If the above $\mathbb{Z}_{2}$-grading preserves the multivector structure of
$\Lambda(V)$, then $\Gamma_{\alpha}$ is a Clifford map for the pair
$(V,g_{\alpha})$, where given $u,v\in V$, the deformed metric $g_{\alpha}$ is
defined by $g_{\alpha}(u,v)=g(u_{0},v_{0})-g(u_{1},v_{1})$, with 
$u_{i}=\pi_{i}(u)$, $v_{i}=\pi_{i}(v)$, $i=0,1$ (see footnote 1).
\label{proposition signature change}
\end{proposition}

\noindent\textbf{Proof.:} As the $\mathbb{Z}_{2}$-grading is assumed to
preserve the multivector structure of $\Lambda(V)$, we have $\alpha(v)\in V$
$\forall v\in V$. This yields 
$(\Gamma_{\alpha}(v))^{2}(x)=(v\wedge+\alpha(v)\lrcorner)
(v\wedge x+\alpha(v)\lrcorner x)=v\wedge(\alpha(v)\lrcorner x)+\alpha(v)\lrcorner(v\wedge x)=
(\alpha(v)\lrcorner v)x$, $\forall x\in V$. Therefore, 
$(\Gamma_{\alpha}(v))^{2}=g(\alpha(v),v)1_{\Lambda(V)}$ and thus
\[
(\Gamma_{\alpha}(v))^{2}=
\genfrac{\{}{.}{0pt}{}
{g(v,v)1_{\Lambda(V)}\textrm{ if }v\in V_{0},}
{-g(v,v)1_{\Lambda(V)}\textrm{ if }v\in V_{1},}
\]
where $V_{i}:=V\cap\mathcal{C}\ell_{i}$, $i=0,1$ (as in the previous
section).$\Box\vspace{10pt}$

Under the conditions above, we can define a Clifford product $\vee_{\alpha}$
in $V^{\wedge}$, associated to $\Gamma_{\alpha}$, by
\[
v\vee_{\alpha}a=v\wedge a+\alpha(v)\lrcorner a, \quad v\in V,a\in V^{\wedge},
\]
extended by linearity and associativity to all of $V^{\wedge}$. It follows that
$(V^{\wedge},\vee_{\alpha})$ is the Clifford algebra associated to $(V,g_{\alpha})$,
where $g_{\alpha}$ is defined in proposition \ref{proposition signature change}.

Given $v\in V$ and $a\in\Lambda^{k}(V)$, with $v_{i}:=\pi_{i}(v)$, this product 
is related to the original Clifford product (denoted by juxtaposition) by 
$v\vee_{\alpha}a=v_{0}\wedge a+v_{0}\lrcorner a+v_{1}\wedge a-v_{1}\lrcorner a=
v_{0}a+(-1)^{k}(a\wedge v_{1}+a\llcorner v_{1})=v_{0}a+\hat{a}v_{1}$. Therefore,
the signature changed product $\vee_{\alpha}$ may be written in terms of the
original one as
\begin{equation}
v\vee_{\alpha}a=v_{0}a+\hat{a}v_{1},\label{new product in terms of the old one}
\end{equation}
where $v\in V$ and $a\in V^{\wedge}$. A more general expression for the
$\vee_{\alpha}$-product between arbitrary multivectors may be obtained from
the above formula by recursion.

Consider now the situation where one wants to change the metric signature from
$(p,q)$ to $(r,s)$, with $p+q=r+s$ 
(see introduction for a discussion on the instances where
this can be useful). To accomplish that, we emulate the Clifford product
associated to this new metric inside the algebraic structure of
$\mathcal{C}\ell_{p,q}$, i.e., using only the algebraic data of $\mathcal{C}\ell_{p,q}$. 
More specifically, suppose that the square of some basis vectors
$e_{i_{1}},\dots,e_{i_{k}}$, of an orthonormal basis 
$\{e_{1},\dots,e_{n}\}\in\mathbb{R}^{p,q}$, are required to change sign in this new 
setting, i.e., when viewed inside the signature changed space. We then define a suitable
$\mathbb{Z}_{2}$-grading for $\mathcal{C}\ell_{p,q}$ by declaring 
$e_{i_{1}},\dots,e_{i_{k}}$ as $\alpha$-odd and the remaining basis vectors as 
$\alpha$-even. In other words, we choose the $\alpha$-parity of the elements in
$\{e_{1},\dots,e_{n}\}$ by
\[
\begin{tabular}{|c|c|c|}
\hline
& $\mathcal{C}\ell_{0}$ & $\mathcal{C}\ell_{1}$\\
\hline
1-vectors & remaining $e_{i}$'s & $e_{i_{1}},\dots,e_{i_{k}}$\\
\hline
\end{tabular}
,
\]
and let this choice generate the $\mathbb{Z}_{2}$-grading in which 
$\alpha$-even ($\alpha$-odd) elements are products of

(a) an even (odd) number of elements in $\{e_{i_{1}},\dots,e_{i_{k}}\}$;

(b) any number of elements in 
$\{e_{1},\dots,e_{n}\}\backslash\{e_{i_{1}},\dots,e_{i_{k}}\}$.

\noindent By the above proposition, the corresponding $\vee_{\alpha}$-product
clearly implements the desired signature change 
$\mathcal{C}\ell_{p,q}\rightarrow\mathcal{C}\ell_{r,s}$. To clarify what is 
going on, we observe that we initially have a space of multivectors 
$V^{\wedge}={\bigoplus\nolimits_{k=0}^{n}}\Lambda^{k}(V)$ (which is not an algebra).
Then, various products can be defined on $V^{\wedge}$. As we have seen,
endowing $V^{\wedge}$ with the exterior product leads to the Grassmann algebra
$\Lambda(V)=(V^{\wedge},\wedge)$, while endowing $V^{\wedge}$ with the
Clifford product leads to the 
Clifford algebra $\mathcal{C}\ell_{p,q}=(V^{\wedge},$Clifford product$)$. In
the same way, the above arguments show that the Clifford algebra associated to
the signature changed metric (with signature $(r,s)$) is given by
$\mathcal{C}\ell_{r,s}=(V^{\wedge},\vee_{\alpha})$. Moreover, the 
$\vee_{\alpha}$-product is parametrized by $\mathbb{Z}_{2}$-gradings and is 
related to the original Clifford product by eq. (\ref{new product in terms of the old one}).

Some examples are in order:

(i) For the \textit{trivial} $\mathbb{Z}_{2}$-grading, where 
$\mathcal{C}\ell_{0}=\mathcal{C}\ell_{p,q}$, i.e.,
\[
\begin{tabular}{|c|c|c|}
\hline
& $\mathcal{C}\ell_{0}$ & $\mathcal{C}\ell_{1}$\\
\hline
1-vectors & $e_{1},\dots,e_{n}$ & ---\\
\hline
\end{tabular}
,
\]
we have $\alpha=$id$_{\mathcal{C}\ell_{p,q}(\mathbb{R})}$ and thus
$\vee_{\alpha}=$[original product]. In other words, the trivial 
$\mathbb{Z}_{2}$-grading yields the trivial signature change (none);

(ii) For the \textit{usual} $\mathbb{Z}_{2}$-grading, i.e.,
\[
\begin{tabular}{|c|c|c|}
\hline
& $\mathcal{C}\ell_{0}$ & $\mathcal{C}\ell_{1}$\\
\hline
1-vectors & --- & $e_{1},\dots,e_{n}$\\
\hline
\end{tabular}
,
\]
we have $\alpha|_{\mathbb{R}^{p,q}}=-$id$_{\mathbb{R}^{p,q}}$ an thus
$\vee_{\alpha}$ yields a change to the opposite metric 
$\mathcal{C}\ell_{p,q}\rightarrow \mathcal{C}\ell_{q,p}$. A straightforward 
calculation shows that given $a,b\in \mathcal{C}\ell_{p,q}$, we have 
$a\vee_{\alpha}b=b_{0}a_{0}+b_{0}a_{1}+b_{1}a_{0}-b_{1}a_{1}$, where 
$a_{i}=\pi_{i}(a)$ and $b_{j}=\pi_{j}(b)$. This is precisely the tilt 
transformation introduced by Lounesto in \cite{Lounesto97}.

(iii) For the $\mathbb{Z}_{2}$-grading
\[
\begin{tabular}{|c|c|c|}
\hline
& $\mathcal{C}\ell_{0}$ & $\mathcal{C}\ell_{1}$\\
\hline
1-vectors & $e_{1},\dots,e_{k-1},e_{k+1},\dots,e_{n}$ & $e_{k}$\\
\hline
\end{tabular}
,
\]
we have $\mathcal{C}\ell_{p,q}\rightarrow \mathcal{C}\ell_{p-1,q+1}$ if $e_{k}$ 
originally squares to $+1$ and $\mathcal{C}\ell_{p,q}\rightarrow \mathcal{C}\ell_{p+1,q-1}$ 
if $e_{k}$ originally squares to $-1$;

(iv) Finally, for the arbitrary $\mathbb{Z}_{2}$-grading
\[
\begin{tabular}{|c|c|c|}
\hline
& $\mathcal{C}\ell_{0}$ & $\mathcal{C}\ell_{1}$\\
\hline
1-vectors & remaining $e_{i}$'s & $e_{i_{1}},\dots,e_{i_{|r-p|}}$\\
\hline
\end{tabular}
,
\]
we have an arbitrary signature change $\mathcal{C}\ell_{p,q}\rightarrow \mathcal{C}\ell_{r,s}$.
Therefore, the product $\vee_{\alpha}$ parametrizes all the possible signature
changes in $\mathcal{C}\ell_{p,q}$ by means of $\mathbb{Z}_{2}$-gradings.

As a final remark, we note that Lounesto's tilt transformation can be
alternatively generalized by the following prescription. Given $a_{i}\in\mathcal{C}\ell_{i}$, 
$b_{j}\in\mathcal{C}\ell_{j}$, we may define 
$a_{i}\vee_{\alpha}^{\prime}b_{j}:=(-1)^{ij}b_{i}a_{j}$, and extend
$\vee_{\alpha}^{\prime}$ as a bilinear product in $V^{\wedge}$. 
A straightforward calculation shows that $\vee_{\alpha}^{\prime}$ is associative and 
preserves the $\mathbb{Z}_{2}$-graded structure of $\mathcal{C}\ell_{p,q}$ in question, 
in the sense that 
$\mathcal{C}\ell_{i}\vee_{\alpha}^{\prime}\mathcal{C}\ell_{j}\subseteq
\mathcal{C}\ell_{i+j(\operatorname{mod}2)}$. By defining convenient 
$\mathbb{Z}_{2}$-gradings exactly as above, we see that $\vee_{\alpha}^{\prime}$ also
provides general signature change maps $C\ell_{p,q}\rightarrow C\ell_{r,s}$.
However, the usual relation between the exterior product and the Clifford
product must be accordingly changed. As a matter of fact, given two 1-vectors
$x,y\in V$, we have $x\wedge y=\frac{1}{2}(xy-yx)$ but 
$x\wedge y=\sum_{ij}(-1)^{ij}\frac{1}{2}(y_{i}\vee_{\alpha}^{\prime}x_{j}-
x_{j}\vee_{\alpha}^{\prime}y_{i})$,
where $x_{i}=\pi_{i}(x)$, $y_{j}=\pi_{j}(y).$ In Lounesto's tilt to the
opposite metric $\mathcal{C}\ell_{1,3}\rightarrow \mathcal{C}\ell_{3,1}$, the latter 
expression simplifies to $\frac{1}{2}(x\vee_{\alpha}^{\prime}y-y\vee_{\alpha}^{\prime}x)$,
but it is easy to see that, in general, this is not the case.

\section{Concluding remarks}

We studied in detail an important class of $\mathbb{Z}_{2}$-graded structures
on a real Clifford algebra $\mathcal{C}\ell(V,g)$. The corresponding
$\mathbb{Z}_{2}$-gradings $\mathcal{C}\ell_{0}\oplus\mathcal{C}\ell_{1}$ are
required to preserve the multivector structure of the underlying Grassmann
algebra over $V$ (see definition \ref{def Z2 pres multivec struc}). A complete
classification for the associated even subalgebras, i.e., for the
$\mathcal{C}\ell_{0}$'s, was obtained. 
As preliminary applications, we first outlined the possibility of
using such general even subalgebras as spinor spaces. After that, we employed
such $\mathbb{Z}_{2}$-graded structures to deform the Clifford product of
$\mathcal{C}\ell(V,g)$, thereby parametrizing all the possible signature
changes on this algebra. This can be useful in signature changing applications
in theoretical physics (see introduction). As we also mentioned at the
introduction, the pervasiveness of $\mathbb{Z}_{2}$-graded structures in
mathematical physics allows us to expect that yet other applications are
likely to be found.

As a last remark, we would like to note that the opposite path to the one
considered here, with a fixed $\mathbb{Z}_{2}$-grading and alternative
multivector structures, has been receiving considerable interest in the
literature. Applications range from models in QFT \cite{Fauser98} to
q-quantization of Clifford algebras \cite{Fauser99} (see also
\cite{FauserAblamowicz00} and references therein).\bigskip

\noindent\textbf{Acknowledgments }\textit{The authors are grateful to B.
Fauser, E. Hoefel, P. Lounesto and W. Rodrigues for useful comments. RAM is
grateful to FAPESP for the financial support (process number 98/16486-8). DM
acknowledges support from the Spanish Ministry of Science and Technology
contract No. BFM2000-0604 and 2000SGR/23 from the DGR of the Generalitat de
Catalunya. JV is grateful to CNPq (300707/93-2) and FAPESP (01/01618-0) for
partial financial support.}

\end{document}